\documentclass[aip,amsmath,amssymb,reprint]{revtex4-1}

\usepackage{graphicx}
\usepackage{dcolumn}
\usepackage{bm}
\begin{document}


\title{Extremely large linear magnetoresistance in Antimony crystal}

\author{Mukesh Kumar Dasoundhi}
\author{Indu Rajput}
\author{Devendra Kumar}
\author{Archana Lakhani}\email{archnalakhani@gmail.com}
\affiliation{UGC-DAE Consortium for Scientific Research, Indore-452001}

\begin{abstract}
In this letter we report the observation of extremely large non-saturating linear magnetoresistance (MR) in Antimony(Sb) crystal. An extremely large magnetoresistance (XMR) of 43000\% at 2K and large unsaturating MR $\sim\ $70\% at room temperature is observed at the magnetic field of 9T. Hall measurements reveal a very high mobility $\sim\ $3.8 x 10$^{4}$ cm$^{2}$/Vs of charge carriers and strong temperature dependence of carrier concentration and mobility. The respective scaling of MR and crossover field (B$_{c}$)  from quadratic to linear MR with mobility and inverse of mobility describes the classical origin of large linear MR in this crystal as suggested by Parish and Littlewood (PL) model for disordered systems.

\end{abstract}
                              
\maketitle

\section{Introduction}

The materials exhibiting magnetic field dependent transport properties are promising candidate for magnetic memory, read access memory and spintronic devices\cite{GMR}. Therefore, the discovery of very large MR in magnetic multilayers, oxide materials and intermetallics have attracted the researchers, particularly the non-magnetic semimetals. The non-magnetic semimetals show variety of exotic phenomena like field induced metal to insulator transition, quantum oscillations, carrier compensation and field induced topological protection from backscattering\cite{MITB,lasb,NaSb2,Wte2N,WTe2flakes}. These exotic behaviors are responsible for XMR in non-magnetic semimetals\cite{lubi,TaSb2,labi2}. Besides this some semimetals (particularly Weyl and Dirac semimetals) possess very high mobility at low temperature and high fields as observed in Cd$_{3}$As$_{2}$ and  Bismuth\cite{Cd3As2,Bi}. This is the reason, semimetals are very good platform to understand the unusual carrier mobility and origin of MR, which has been observed in various emerging materials of technological use. Bismuth and WTe$_2$ exhibits high MR in which electron(e) - hole(h) pockets coexist at fermi surface. In some semimetals the conduction band and valence band touch the fermi surface at same point in momentum space (k-space)\cite{low}. Those  semimetals exhibit very high carrier mobility and large linear field dependent MR  such as  Ag$_{2+\delta}$Se, Ag$_{2+\delta}$Te and Cd$_{3}$As$_{2}$\cite{Ag2Se,Cd3As2}. The large linear magnetoresistance (LMR) is also found in highly disordered  non-magnetic materials such as  Bi films, epitaxial graphene, topological insulators\cite{Bifilms,egraphene,graphene,WTe2PRL,narayanan,NbP,Bi2Se3}. Abrikosov proposed that LMR has quantum origin in materials with linear dispersion of energy at band touching points, arising when electrons are filled in first Landau levels(LL)\cite{AA}. Whereas Parish and Littlewood (PL) assumed LMR has a classical origin and explained LMR in  Ag$_{2+\delta}$Se and Ag$_{2+\delta}$Te by modeling the materials as random resistor network due to disorder- induced mobility fluctuations\cite{PL}. The inhomogeneities are generally associated with the action of galvanomagnetic effects due to reduction of carrier mobility caused by impurity scattering. In strong disordered systems where length scale of inhomogeneities is of the order of mean free path, the observed MR is positive and linear with applied field in presence of spacial fluctuation of resistivity\cite{PL,PL2,narayanan}. This provides an opportunity opportunity to design the materials and devices where LMR particularly originates from  inhomogeneities or strong disorder and MR can be tuned linearly with applied field.\par
In this letter we report the magnetotransport study of Antimony crystal at low temperature and high fields. We have observed extremely large non-saturating MR upto 9T at low temperatures. The observed MR is linear in nature at higher fields (above 4.6T). This type of dependency of MR on magnetic field without saturation make the compounds suitable for development of magnetoresistive devices such as magnetic sensors and switches. In order to know the origin of linear MR, the carrier mobility and carrier concentrations are estimated. The observed linear MR scales with carrier mobility and inverse of carrier concentration, which confirms the classical origin of linear MR that depends on inhomogeneity or disorder present in the crystal. 
\section{Experimental}
Oriented crystal of Antimony was grown by controlled heating and cooling of Sb granules (5N purity) in an evacuated atmosphere\cite{ICAM}. The phase purity and crystal structure was determined by X-ray Diffraction(XRD) using Bruker D8 Advance Diffractometer with Cu-K$\alpha$ radiation. To get the lattice parameters, Le Bail fitting of powder XRD has been done using FULLPROF software. All the electrical transport measurements are carried out  in Physical Property Measurement System from 2K to 300K and in applied field upto 9T. The standard 4-probe and 5-probe method was used for resistivity and Hall measurements respectively.

\section{Results and Discussions}
\subsection{Structural Analysis}
Figure 1(a) shows the powder XRD pattern along with Le Bail fitting of crushed Sb crystal. The obtained Bragg reflections are well indexed with space group (166) as reported in JCPDF. The obtained lattice parameters are a=b=4.30\AA, c=11.27\AA, and $\alpha$=$\beta$=90$^{0}$, $\gamma$=120$^{0}$. Figure 1(b) shows the XRD pattern of cleaved crystal indicating preferred orientation along  $<003>$ direction with a slight contribution of (202) plane. The inset of Fig.1(b) shows the Williamson-Hall (W-H) plot of crystal. The microstrain present in the crystal is $\sim\ $4 x 10$^{-4}$ and the crystallite size is $\sim\ $3$ \mu $m. The relatively large value of microstrain suggests the presence of local strain field in the crystal. The positive value of microstrain indicates the dominance of local compressive strain field arising from lattice disorder other than vacancies such as the grain boundaries, dislocation etc.\cite{Strain,berry}.  

\begin{figure}[hbtp]
\centering
\includegraphics[width=0.8\linewidth]{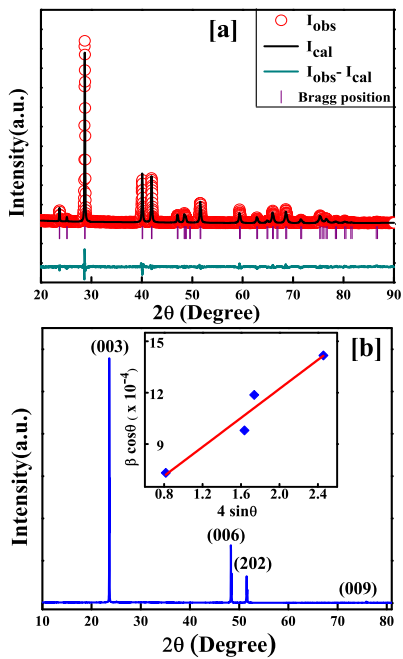}
\caption{(a) Powder XRD of Sb with Le Bail fitting. Red circles represent the experimental data, black solid  line is the Le Bail fitting for (166) space-group, violet bars show the Bragg peak positions, and solid dark cyan line is the difference between experimental and fitted data. (b) XRD pattern of cleaved crystal. Inset shows the Williamson-Hall plot of crystal.}
\end{figure}

 \subsection{Resistivity and Magnetoresistance Analysis} 
    Magnetoresistance of a material at a given temperature is calculated as, MR=[$ \rho $$_x$$_x$(H)-$ \rho $$_x$$_x$(0)/$ \rho $$_x$$_x$(0)]x100, where $ \rho $$_x$$_x$(H) is the resistivity at applied field H and $ \rho $$_x$$_x$(0) is the zero field resistivity. Figure 2(a) shows the MR measured at various temperatures  in magnetic field up to 9T.  The MR measured at 2K and 9T is about 43000\%, which is greater than the XMR reported in rare-earth monopnictide materials like GdSb, LaBi, etc. \cite{gdsb,labi}. This MR is non-saturating upto the measured field as shown in fig.2(a). Figure 2(b) displays the variation of MR at 9T with temperature. The MR decreases gradually from 2K to 15K and thereafter a drastic decrease of MR is seen upto 100K and MR reduces to 70\% at room temperature. This room temperature MR is comparatively much higher than various XMR materials like GdSb(60\%), LuBi(6\%), YBi(7\%)\cite{gdsb,lubi}. The large MR observed in Antimony at room temperature, makes it a promising candidate for magnetic memory and spintronic devices. In general, the XMR in semimetals is associated with compensation of charge carriers and their high mobility at low temperature and high fields\cite{gdsb,lasb,NaSb2,lubi,WTe2PRL,Cd3As2}.
    
    \begin{figure}[hbtp]
    	\centering
    	\includegraphics[width=1\linewidth,height=0.9\linewidth]{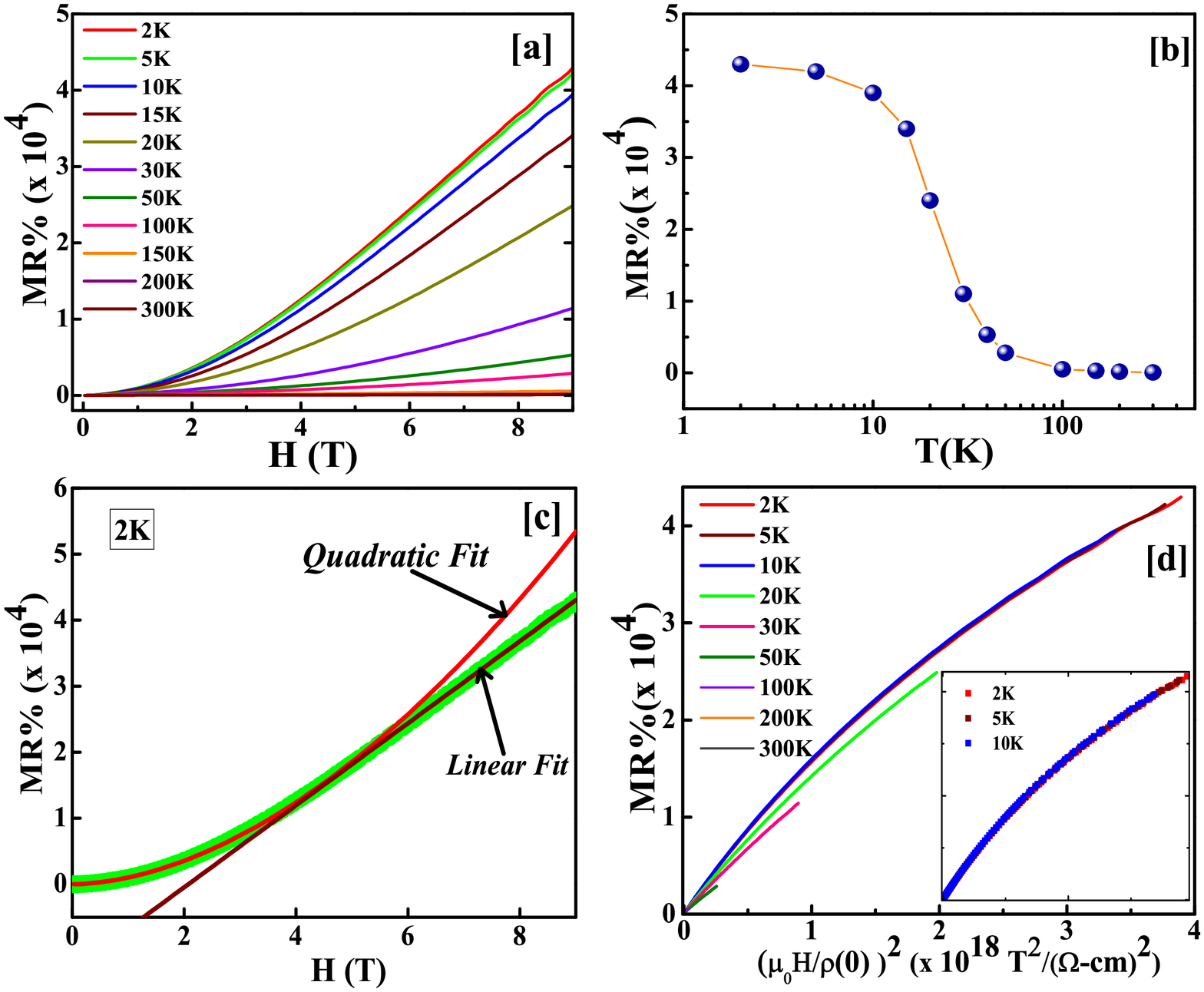}
    	\caption{(a) Variation of MR with field at various temperatures. (b) Temperature dependence of MR at 9T. (c) MR data at 2K, showing quadratic fit (red line) at lower fields and linear fit (brown line) at higher fields. (d) MR at various temperatures are scaled using Kohler's rule. Inset shows the Kohler's scaling of MR at 2K, 5K and 10K.}
    \end{figure}
    
     Figure 2(c) shows the MR with field  at 2K, representing the crossover from quadratic to linear dependence on magnetic field.  We find that the MR upto 4.6T follows quadratic behavior and then at higher fields it follows linear behaviour, similar behavior is reported by Devendra et al. in Bi$_{2}$Se$_{3}$ single crystal and Liang et al. in Cd$_{3}$As$_{2}$ crystal\cite{Bi2Se3,Cd3As2}. The crossover field, B$_{c}$ is obtained by extracting the field where the quadratic dependence of low field MR and linear dependence of high field MR curve crosses. The linear behaviour of MR at high fields may have a quantum or classical origin\cite{AA,PL}, which is discussed in the next section. Figure 2(d) shows the Kohler scaling of Antimony with field at various temperatures. As per Kohler's rule, for single type of charge carriers with constant temperature dependent scattering time at all points of Fermi surface, the MR curves at various temperatures should follow the scaling $\Delta\rho $/$\rho $(0)=f(H/$ \rho $(0)) and fall on a single curve. Inset shows the Kohler's scaling of MR measured at 2K, 5K and 10K which collapse into a single curve. This suggests the presence of same type of charge carrier with unique temperature dependent scattering rate at low temperatures. But at temperatures above 10K, the deviation from Kohler's rule is observed as shown in fig. 2(d) where MR curves do not fall into a single curve. This violates the Kohler's rule and  suggests the presence of more than one type of charge carriers having different temperature dependent mobility, which is further examined by Hall resistivity measurements.

\subsection{Hall effect}
For better understanding of carrier transport, Hall resistivity measurements are performed. The field dependence of Hall resistivity at high fields exhibits linear behaviour whereas at low fields it is nonlinear. The nonlinear behaviour of Hall resistivity at low fields further suggests the involvement of more than one type of charge carriers as indicated by Kohler's scaling above 10K. For simplicity, we have adopted single carrier Drude model for the estimation of carrier concentration and mobility at higher fields because the present work discusses the transport behaviour at higher fields. The carrier concentration is calculated by n=1/eR$_{H}$, where e is charge of electron and R$_{H}$ is Hall coefficient. The Hall mobility is calculated as $\mu$= R$_{H}$/$ \rho $$_x$$_x$(H=0). 

 \begin{figure}[hbtp]
	\centering
	\includegraphics[width=0.8\linewidth]{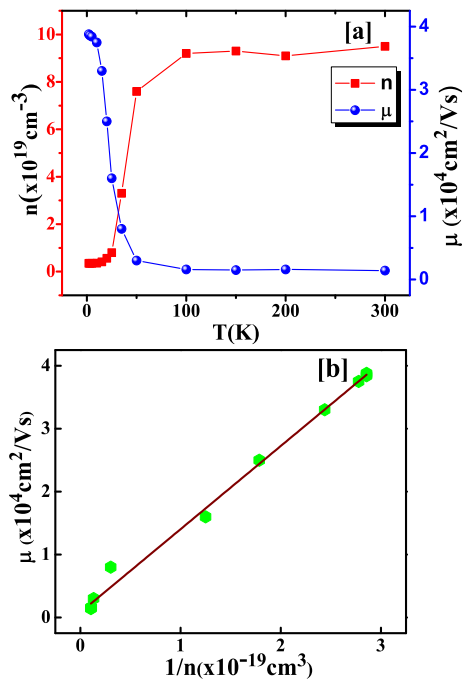}
	\caption{(a) Variation of n and $\mu$ with temperature. (b) Plot of $\mu$ and 1/n at corresponding temperatures. Solid (brown) line is linear fit to data.}
\end{figure}

Figure 3(a) shows the variation of carrier concentration and mobility with temperature. Figure 3(b) shows the variation of mobility with inverse of carrier concentration at corresponding temperatures. It is clearly seen that mobility scales linearly with 1/n i.e. ($\mu\propto$1/n). The carrier concentration and mobility is found to be $\sim\ $3.5 x 10$^{18}$cm$^{-3}$ and $\sim\ $3.8 x 10$^{4}$ cm$^{2}$/Vs at 2K and 9T respectively. The values of n and $\mu$ are close to the transition metal dipnictide TaSb$_2 $ reported by Sudesh et al. and less than Dirac semimetal Cd$_{3}$As$_{2}$ reported by Liang et al.\cite{Cd3As2,TaSb2}. The high mobility of charge carriers can be observed when there is a zero band gap or linear dispersion of bands near the Fermi level where the charge carriers experience a very low effective mass or behave like  massless particles\cite{low,YPtSb,Phonon-glass}. Linear dispersion of surface states in Antimony has been reported by Narayan et al. using ab initio calculation and ARPES simulations\cite{Narayan} and Fauque et al. have also observed XMR in Sb single crystal which is attributed to field dependent mobility\cite{Benoit}. Therefore, observation of XMR in case of Sb could be due to high carrier mobility at low temperature and high fields and the ultra-high mobility of carrier might be arising from linear dispersion of bands. In our case the linear dispersion of bands in Antimony could be due to the presence of small energy gap and Fermi level lying near the band gap\cite{SbZt,YPtSb,low}. 
 
\subsection{Origin of large linear MR}

 The non-saturating linear MR in a material may have a classical or quantum origin. The classical effect is related to the inhomogeneity and disorder present in the system whereas the quantum effect is related to linear dispersion of energy at band touching points\cite{PL,AA}. According to Herring et al., Large linear MR  may arise if there are small fluctuations in local conductivity in the limit of weak disorder\cite{Herring}. According to Parrish and Littlewood (PL) model,  linear MR arises in strong disordered systems because the path of current is distorted  due to inhomogeneous distribution of carrier concentration(n) and mobility ($\mu$)\cite{PL}. 
 
 \begin{figure}[h!]
 	\centering
 	\includegraphics[width=1\linewidth,height=0.9\linewidth]{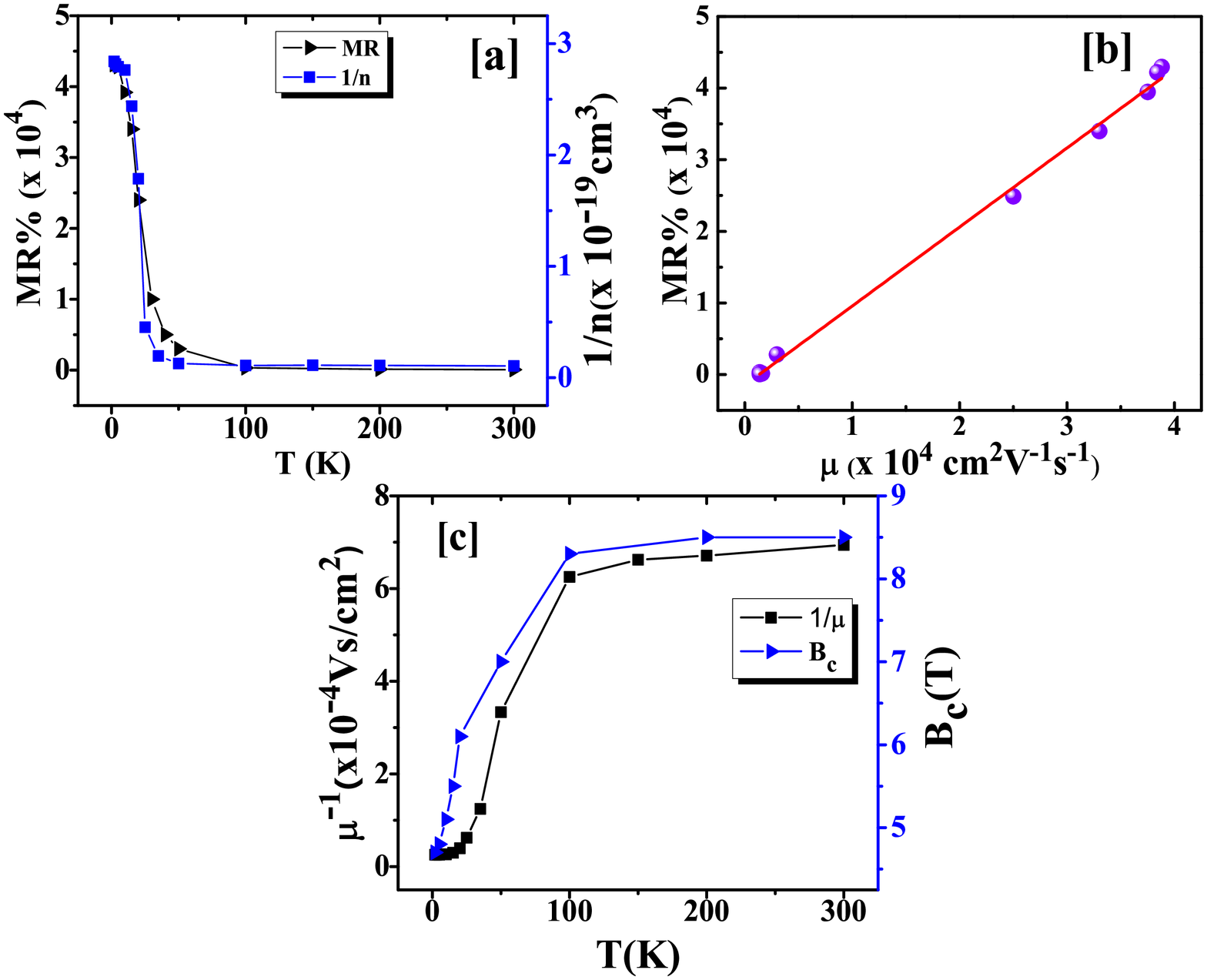}
 	\caption{(a) Temperature dependence of MR and 1/n at 9T. (b) MR (at 9T) with carrier mobility at respective temperatures. The red solid line is the linear fit to data. (c) Temperature dependence of crossover field (B$_{c}$) and inverse of mobility($\mu$).}
 \end{figure}

 They modeled a 3-dimensional random resistor network to discover the field dependence of MR in inhomogeneous systems, where regions of different local carrier concentration and mobility $\mu$ are represented by a 4-terminal resistor.  The linear MR originating from this process is associated with average mobility $<\mu>$ and width of mobility disorder $\Delta\mu $. The LMR follows MR$ \propto\mu $, B$_{c}\propto\mu^{-1}$, when $\Delta\mu$/$\mu<$1 and MR$ \propto\Delta\mu $,  B$_{c}\propto \Delta\mu^{-1}$ when $\Delta\mu$/$\mu>$1. Figure 4(a) displays the variation of MR and 1/n with temperature for our crystal. The values of MR and 1/n follow similar trend with temperature i.e. MR and carrier concentration are inversely related to each other. The drastic change of carrier concentration and mobility with temperature suggests the presence of disorders or defects in the system. Figure 4(b) shows the variation of MR with mobility at respective temperatures. It is clearly seen that the linear MR $\propto\mu$, in the entire range of measured temperatures. Figure 4(c) shows the temperature dependence of crossover field B$_{c}$ and inverse of carrier mobility $\mu^{-1}$. B$_{c}$ increases with temperature whereas mobility decreases with temperature with B$_{c}$ following the same trend as $\mu^{-1}$. Hence, the linear MR $\propto\mu$ and B$_{c}\propto\mu ^{-1}$ i.e. crossover field scales with inverse of mobility. This satisfies the PL model for narrow mobility distribution, that is $\Delta\mu$/$\mu <$1 and confirms the classical origin of LMR in our crystal. This type of linear MR following the PL model is also shown by Ag$_{2+\delta}$Se and Ag$_{2+\delta}$Te, graphene, Cd$_{3}$As$_{2}$, Bi$_{2}$Se$_{3}$ crystals\cite{Ag2Se,PL2,egraphene,graphene,Cd3As2,Bi2Se3}.\par
From W-H plot, it is clear that there is microstrain in our crystal suggesting the presence of lattice disorder.The lattice disorder could be in the form of stacking faults and grain boundaries between the crystallites which have the length scale of crystallite size i.e. $ \sim $3$ \mu $m. The mean free path of charge carriers estimated from the carrier density and mobility is  $\sim\ $1.2$\mu$m. The similiar order of lattice disorder and mean free path indicates that these crystalline defects act as the source of inhomogeneities from where the carriers experience multiple scattering. These crystalline disorders causes fluctuation in carrier mobility responsible for linear MR. 
 
\section{Conclusion}
In conclusion, we have observed and investigated the extremely large unsaturated  linear magnetoresistance (43000\%) at low temperature as well as at room temperature (70\%) upto 9T magnetic field. Kohler's scaling suggests the strong temperature dependence of carrier concentration and mobility of charge carriers, which is well corroborated by Hall resistivity measurements. A crossover from quadratic to linear MR behavior is seen  at B$_{c}$, which increases with increasing temperature. The MR and B$_{c}$ scales with mobility satisfying the conditions of PL model, confirming the classical origin of linear MR due to crystalline disorder and defects present in our crystal. The accomplishment of huge magnetoresistance in the low cost semimetal element like Sb could play a vital role in the realization of magnetic memory devices and magnetic sensors for hard disks.

\section{Acknowledgment}
We thank M. Gupta and L. Behera for XRD measurements.

\end{document}